\documentclass[aps,amsmath,twocolumn,amssymb,floatfix,showpacs,superscriptaddress,nofootinbib,longbibliography]{revtex4-1}
\usepackage{braket}
\usepackage[dvipsnames]{xcolor}
\usepackage{float}
\usepackage{subfigure}
\usepackage[colorlinks=true,linktoc=page,linkcolor=blue,citecolor=blue,urlcolor=blue]{hyperref}

\newcommand{\vect}[1]{\boldsymbol{\mathrm{#1}}}
\mathchardef\mhyphen="2D 

\newcommand{\ie}{{\it i.e.,\,\,}}

\newcommand\bea{\begin{eqnarray}}
\newcommand\eea{\end{eqnarray}}
\newcommand\beq{\begin{equation}}  
\newcommand\eeq{\end{equation}}

\newcommand{\non}{\nonumber}  
\usepackage[normalem]{ulem}
\definecolor{lime}{HTML}{A6CE39}
\usepackage{sidecap,tikz}
\DeclareRobustCommand{\orcidicon}{\hspace{-1.0mm}
	\begin{tikzpicture}
		\draw[lime, fill=lime] (0.0,0.0) 
		circle [radius=0.15] 
		node[white] {{\fontfamily{qag}\selectfont \tiny \,ID}};
		\draw[white, fill=white] (-0.0525,0.095) 
		circle [radius=0.007];
	\end{tikzpicture}
	\hspace{-3.0mm}
}
\foreach \x in {A, ..., Z}{\expandafter\xdef\csname orcid\x\endcsname{\noexpand\href{https://orcid.org/\csname orcidauthor\x\endcsname}
		{\noexpand\orcidicon}}
}

\AtBeginDocument{%
	\newwrite\bibnotes
	\def\bibnotesext{Notes.bib}
	\immediate\openout\bibnotes=\jobname\bibnotesext
	\immediate\write\bibnotes{@CONTROL{REVTEX41Control}}
	\immediate\write\bibnotes{@CONTROL{%
			apsrev41Control,author="08",editor="1",pages="1",title="1",year="1"}}
	\if@filesw
	\immediate\write\@auxout{\string\citation{apsrev41Control}}%
	\fi
}%
\begin{document}


\title{Non-Hermitian higher-order topological superconductors in two-dimension: statics and dynamics}

\author{Arnob Kumar Ghosh\orcidA{}}
\email{arnob@iopb.res.in}
\affiliation{Institute of Physics, Sachivalaya Marg, Bhubaneswar-751005, India}
\affiliation{Homi Bhabha National Institute, Training School Complex, Anushakti Nagar, Mumbai 400094, India}
\author{Tanay Nag\orcidB{}}
\email{tanay.nag@physics.uu.se}
\affiliation{Department of Physics and Astronomy, Uppsala University, Box 516, 75120 Uppsala, Sweden}

\begin{abstract}
	Being motivated by intriguing phenomena such as the breakdown of conventional bulk boundary correspondence and emergence of skin modes in the context of non-Hermitian (NH) topological insulators, we here propose a NH second-order topological superconductor (SOTSC) model that hosts Majorana zero modes (MZMs). Employing the non-Bloch form of NH Hamiltonian, we topologically characterize the above modes by biorthogonal nested polarization and resolve the apparent breakdown of the bulk boundary correspondence. Unlike the Hermitian SOTSC, we notice that the MZMs inhabit only one corner out of four in the two-dimensional NH SOTSC. Such localization profile
	of MZMs is protected by mirror rotation symmetry  and
	remains robust under on-site random disorder. We extend the static MZMs into the realm of Floquet drive. We find anomalous $\pi$-mode following low-frequency mass-kick in addition to the regular $0$-mode that is usually engineered in a high-frequency regime. We further characterize the regular $0$-mode with biorthogonal Floquet nested polarization. Our proposal is not limited to the $d$-wave superconductivity only and can be realized in the experiment with strongly correlated optical lattice platforms. 
\end{abstract}

\maketitle
\textcolor{blue}{\textit{Introduction.---}} In recent times, topological phases in insulators and superconductors are extensively studied theoretically~\cite{hasan2010colloquium,Haldane88,Kane05,bernevig2006quantum,alicea2012new,beenakker2013search} as well as experimentally~\cite{jotzu2014experimental,das2012zero}. The conventional bulk boundary correspondence (BBC) for first-order topological phase is generalized for $n(>1)$th-order topological insulator (TI)~\cite{benalcazar2017,benalcazarprb2017,Song2017,Langbehn2017,schindler2018,Franca2018,wang2018higher,Roy2019,Szumniak2020,Ni2020,BiyeXie2021,saha2022dipolar} and  topological superconductor~\cite{Zhu2018,Liu2018,Yan2018,WangWeak2018,Zhang2019,ZhangFe2019PRL,Volpez2019,YanPRB2019,Ghorashi2019,Wu2020,jelena2020HOTSC,BitanTSC2020,SongboPRR22020,kheirkhah2020vortex,YanPRL2019,AhnPRL2020,luo2021higherorder2021,QWang2018,Ghosh2021PRB,RoyPRBL2021,li2021higher} in $d\ge2$ dimensions where there exist  $n_c=(d-n)$-dimensional boundary modes. The zero-dimensional (0D) corner and one-dimensional (1D) hinge  modes are thus the hallmark signatures of higher-order topological insulator (HOTI) and  higher-order topological superconductor (HOTSC). The dynamic analog of these phases are extensively studied for Floquet HOTI (FHOTI)~\cite{Bomantara2019,Nag19,YangPRL2019,Seshadri2019,Martin2019,Ghosh2020,Huang2020,HuPRL2020,YangPRR2020,Nag2020,ZhangYang2020,bhat2020equilibrium,GongPRBL2021,JiabinYu2021,Vu2021,ghosh2022systematic,du2021weyl,ning2022tailoring} and Floquet HOTSC (FHOTSC)~\cite{PlekhanovPRR2019,BomantaraPRB2020,RWBomantaraPRR2020,RWBomantaraPRR2020,ghosh2020floquet,ghosh2020floquet2,VuPRBL2021,ghosh2022dynamical}.

The realm of topological quantum matter is transcended from the Hermitian system to the non-Hermitian (NH) system due to the practical realization of TI phases in meta-materials~\cite{parappurath2020direct,yang2019realization,Malzard15,regensburger2012parity} where energy conservation no longer holds~\cite{el2018non,Denner2021}. The NH description has a wide range of applications, including systems with source and drain~\cite{Musslimani08,Makris08}, in contact with the environment~\cite{Bergholtz19,Yang21,San-Jose2016}, and involving quasiparticles of finite lifetime~\cite{kozii2017non,Yoshida18,Shen18}. Apart from the complex eigenenergies and non-orthogonal eigenstates, the NH Hamiltonians uncover a plethora of intriguing phenomena in TI~\cite{YaoPRLSecond2018,KawabataPRX2019,Bergholtz2021,Sone2020,Denner2021} that do not have any Hermitian analog. For instance, NH Hamiltonian becomes non-diagonalizable at exceptional points (EPs) where eigenstates, corresponding to degenerate bands, coalesce~\cite{bender2007making,heiss2012physics}; line and point are two different types of gaps in these systems that can be adiabatically transformed into a Hermitian and NH systems, respectively~\cite{KawabataPRX2019}; the conventional Bloch wave functions do not precisely indicate the topological phase transitions under the open-boundary conditions (OBCs) leading to the breakdown of the BBC  \cite{YaoPRL2018,Kunst18,helbig2020generalized,Borgnia20,Koch2020,ZirnsteinPRL2021,TakaneJPS2021}; consequently, the  non-Bloch-wave behavior results in the skin effect where the bulk states accumulate at the boundary \cite{YaoPRL2018,Kunst18,helbig2020generalized,Kawabata20b}, and the structure of topological invariants become intricate \cite{YaoPRLSecond2018,Kai20,Gong18,Yin18}. The EPs are studied in the context of Floquet NH Weyl semimetals \cite{Banerjee20,Chowdhury21}.

While much has been explored on the HOTI phases in the context of NH systems   \cite{NoriNHPRL2019,LuoGainLossPRL2019,EdvardssonNHPRB2019,ZhiwangPRL2019,GongNonreciprocalPRL2019,Ya_JiePRA2020,OkugawaPRB2020,OkugawaPRB2021,Shiozaki2021,YangPRBL2021}, HOTSC counterpart, along with its dynamic signature, is yet to be examined. Note that NH 1D nanowire with $s$-wave pairing and $p$-wave SC chain are studied for the Majorana zero modes (MZMs)       \cite{OkumaPRL2019,LieuPRB2019,Avila2019,ZhaoPRB2021,WangPRB2021,liu2021nonhermiticity,ZhouNHFloquet1DSCPRB2020}. We, therefore, seek the answers to the following  questions that have not been addressed so far in the context of proximity induced HOTSC with non-hermiticity
-(a) How does the BBC change as compared to the Hermitian case? (b) Can one use the concept of biorthogonal nested-Wilson-loop to characterize the MZMs there similar to that for HOT electronic modes \cite{LuoGainLossPRL2019}? (c) How can one engineer the anomalous FHOTSC phase for the NH case?

\begin{figure}[]
	\centering
	\subfigure{\includegraphics[width=0.47\textwidth]{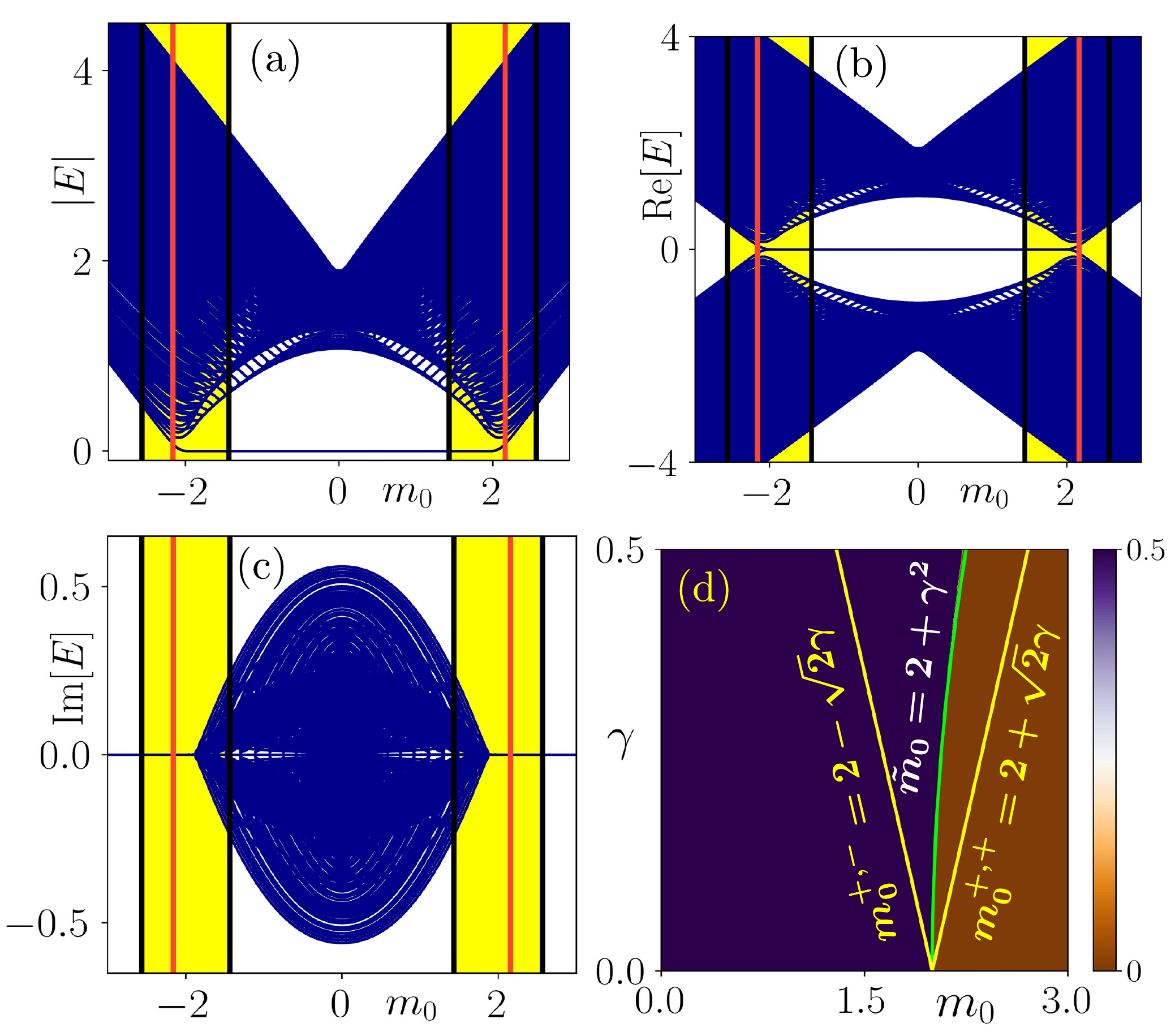}}
	\caption{We show $\lvert E \rvert$, ${\rm Re}[E]$, and ${\rm Im}[E]$, obtained from real space Hamiltonian under OBC in all directions using Eq.~(\ref{eq1_TSC}),  as a function of $m_0$ in (a), (b), and (c), respectively. The mid-gap MZMs disappear into the bulk bands at $m_0={\tilde m_0}=\pm (t_x+t_y+{\gamma_x^2}/{2\lambda_x^2}+{\gamma_y^2}/{2\lambda_y^2})$, defined by the red lines. The EPs $m_0^{s,\pm}= s(t_x+t_y)\pm\sqrt{\gamma_x^2+\gamma_y^2} $ with $s=\pm$ are marked by black lines within which ${\rm Re}[E(k)]$ associated with $\mathcal{H}(\vect{k})$ remains gapless as designated by yellow-shaded region.  (d) The topological phase diagram is depicted in the $m_0 \mhyphen \gamma$ plane using nested polarization $\langle \nu_{y, \mu'}^{\nu_{x}}\rangle$ Eq.~(\ref{npola}). The yellow and green lines correspond to $m_0^{+,\pm}$ and ${\tilde m_0}$, respectively while the later separates SOTSC phase $\langle \nu_{y, \mu'}^{\nu_{x}}\rangle=0.5$ from the trivial phase $\langle \nu_{y, \mu'}^{\nu_{x}}\rangle=0.0$. The parameters used here are  $t_x=t_y=\lambda_x=\lambda_y=\Delta=1.0$ and $\gamma_x=\gamma_y=0.4$.
	}
	\label{Fig:sotscloop}
\end{figure}
\begin{figure}[]
	\centering
	\subfigure{\includegraphics[width=0.49\textwidth]{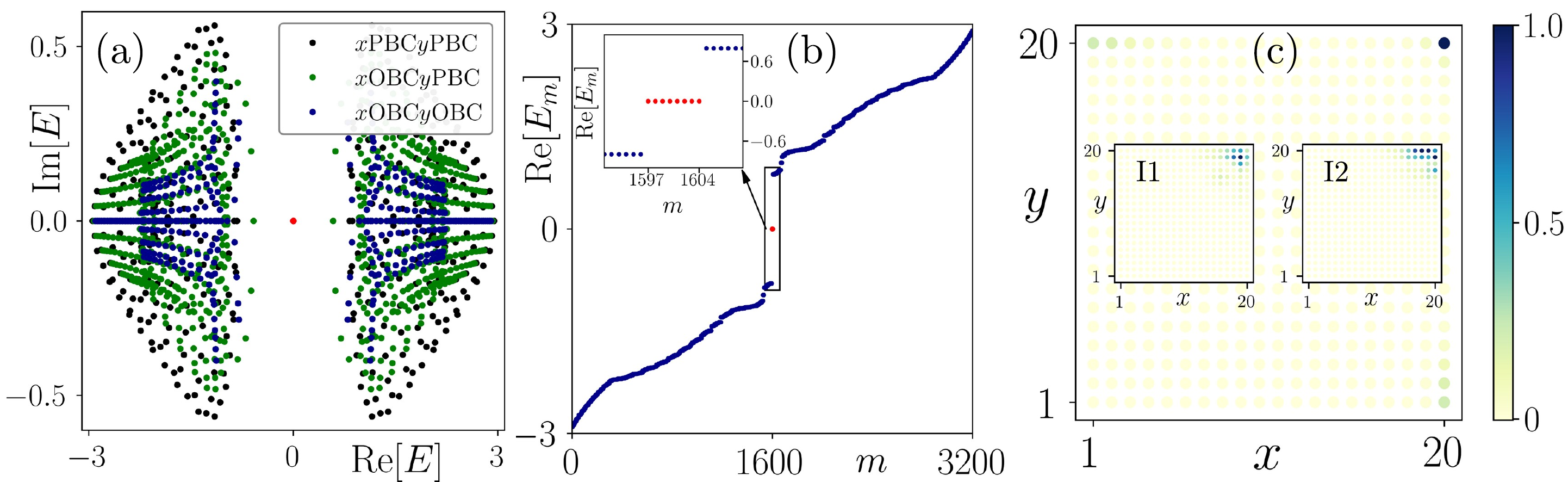}}
	\caption{(a) The  eigenvalue spectrum 
	for the real space 2D system Eq.~(\ref{eq1_TSC}), obeying PBC in both direction (black dots), PBC in $y$ and OBC in $x$-direction (green dots), and OBC in both directions (blue dots) are depicted in complex energy plane.
	The zero-energy mode, obtained from OBC, is marked by red dots. (b) ${\rm Re}[E_m]$ as a function of the state index $m$ is displayed  where eight mid-gap MZMs are highlighted in the inset. (c) The LDOS, associated with eight MZMs in (b),   show sharp localization only at one corner. The LDOS for typical bulk states are shown in insets I1 (for $E_m=-2.631839$) and I2 (for $E_m=-1.738466+0.130006i$). We use $m_0=1.0$, while other parameters are same as Fig.~\ref{Fig:sotscloop}.
	}
	\label{Fig:sotsc}
\end{figure}
\begin{figure}[]
	\centering
	\subfigure{\includegraphics[width=0.49\textwidth]{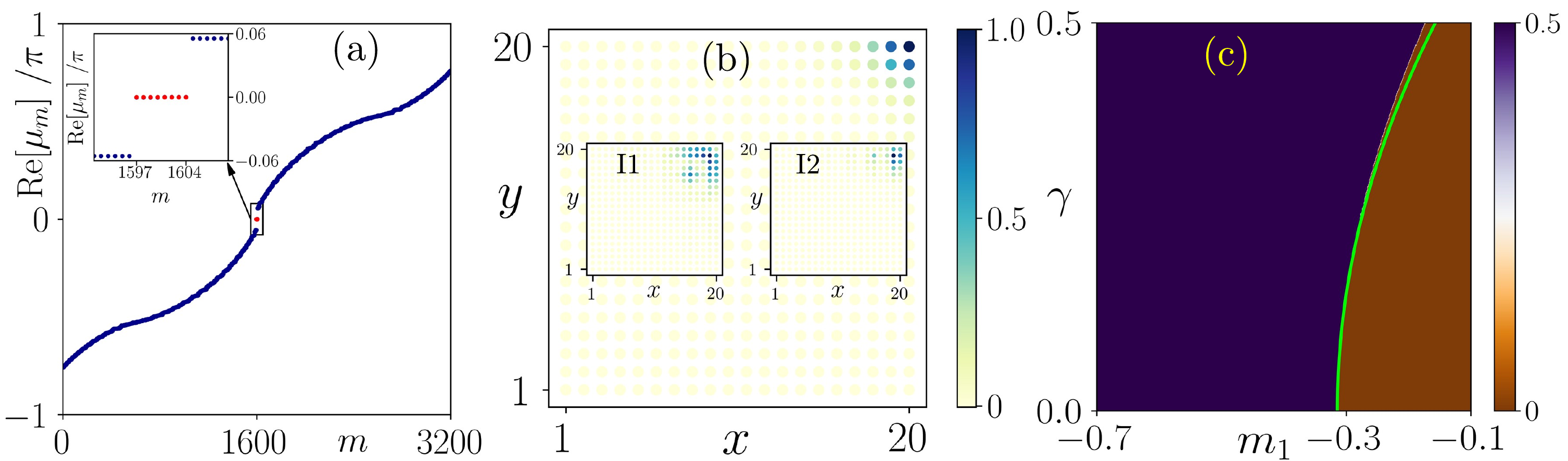}}
	\caption{(a) The real part of the quasi-energy spectrum $E_m$, obtained from Eq.~(\ref{Foperator}) under OBC, are shown with eight Floquet Majorana $0$-modes in the inset. (b) The LDOS, associated with eight MZMs in (a), exhibits substantial localization only at one corner similar to Fig.~\ref{Fig:sotsc} (c). The 
		LDOS for typical bulk modes with $E_m=-0.697702\pi$ and $-0.453177 \pi$ are demonstrated in insets I1 and I2. (c) The  topological phase diagram is depicted in the $m_1\mhyphen \gamma$ plane where the Floquet SOTSC phase is characterized by the  average Floquet nested Wannier sector polarization $\langle \nu_{y, \mu'}^{{\rm F},\nu_{x}}\rangle=0.5$ (blue region). The phase boundary, marked by green line, is consistent with Eq.~(\ref{modifiedmassFloquet}).    
		We consider $m_0=2.5,~m_1=-0.4,~\Omega=10.0$ such that we start from the trivial phase deep inside the brown region in Fig.~\ref{Fig:sotscloop} (c).
	}
	\label{Fig:Floquetsotsc}
\end{figure}
\begin{figure}[]
	\centering
	\subfigure{\includegraphics[width=0.49\textwidth]{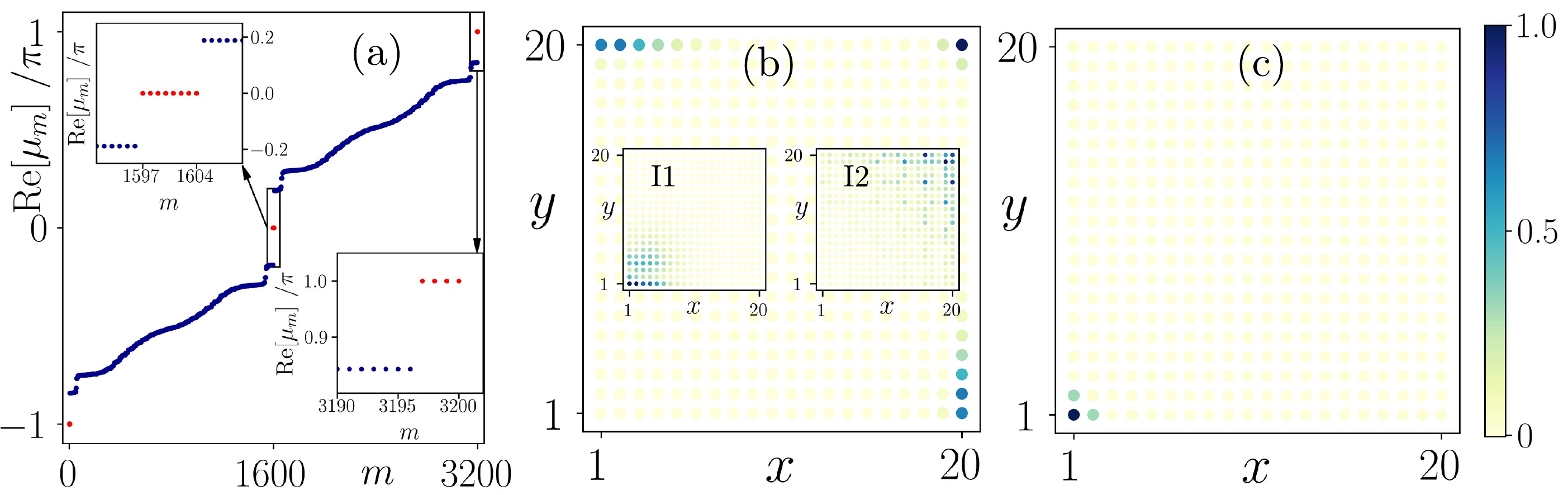}}
	\caption{(a) We repeat Fig.~\ref{Fig:Floquetsotsc} (a) considering the low frequency mass-kick with $\Omega=5.0$ where we find eight regular  $0$- and anomalous $\pi$-modes simultaneously. We show the LDOS  associated with ${\rm Re}[E_m]=0$- and ${\rm Re}[E_m]=+\pi$ in (b) and (c), respectively. 
	The LDOS for Floquet bulk modes  with $E_m=-0.838779 \pi - 0.085397\pi i$ and $-0.471932 \pi$ are respectively depicted in the insets I1 and I2 of (b). We consider $m_0=0,m_1=1.5, \Omega=5.0$. 
	}
	\label{Fig:anomaloussotsc}
\end{figure}

Considering the NH TI in the proximity to a $d$-wave superconductor, we illustrate the generation of the NH second-order topological superconductor (SOTSC). The breakdown of BBC is resolved with the non-Bloch nature of the NH Hamiltonian, where phase boundaries, obtained under different boundary conditions, become concurrent with each other (see Fig.~\ref{Fig:sotscloop}). The SOTSC phase is characterized by the non-Bloch nested polarization. We demonstrate the NH skin effect where MZMs and bulk modes both display substantial corner localization (see Fig.~\ref{Fig:sotsc}). We further engineer the regular $0$- and anomalous $\pi$-mode employing the mass-drive in high- and low-frequency regimes, respectively (see Figs.~ \ref{Fig:Floquetsotsc} and \ref{Fig:anomaloussotsc}). We characterize the regular dynamic $0$-mode by the non-Bloch Floquet nested polarization.

\textcolor{blue}{\textit{Realization of NH SOTSC.---}} We contemplate the following Hamiltonian of the NH SOTSC, consisting of NH TI $H_{\rm TI} (\vect {k})$ and $d$-wave proximitized superconductivity ~\cite{Yan2018,ghosh2020floquet2}
\begin{equation}
	\mathcal{H}(\vect{k})=
	\begin{pmatrix}
		H_{\rm TI} (\vect {k})-\mu & \Delta \\
		\Delta^* & \mu-\tilde{H}_{\rm TI} (-\vect {k})\   \\
	\end{pmatrix}\ ,
	\label{eq1_TSC}
\end{equation}
where, $\tilde{H}_{\rm TI} (\vect {k})=U^{-1}_{\mathcal{T}} H^*_{\rm TI} (\vect {k}) U_{\mathcal{T}}$. Here, $H_{\rm TI} (\vect {k})=(\lambda_x \sin k_x +i \gamma_x) \sigma_x s_z + (\lambda_y \sin k_y +i \gamma_y) \sigma_y s_0 + (m_0 -t_x \cos k_x - t_y \cos k_y ) \sigma_z s_0=H^{\rm H}_{\rm TI} (\vect {k}) + i \gamma_x \sigma_x s_z + i \gamma_y \sigma_y s_0$, that preserves ramified (time-reversal symmetry)~TRS:  $U_{\mathcal{T}} \mathcal{H}^*_{\rm TI}(\vect{k}) U^{-1}_{\mathcal{T}}=\mathcal{H}_{\rm TI}(-\vect{k})$ and  (particle-hole symmetry)~PHS$^\dagger$: $U_{\mathcal{C}} \mathcal{H}^*_{\rm TI}(\vect{k}) U^{-1}_{\mathcal{C}}=-\mathcal{H}_{\rm TI}(-\vect{k})$ with $U_{\mathcal{T}}=\sigma_0 s_y$ and $U_{\mathcal{C}}=\sigma_x s_0$, respectively~\cite{KawabataPRX2019}. The $d$-wave superconducting paring is given by $\Delta(\vect{k})=\Delta (\cos k_x - \cos k_y)$; whrereas $\gamma_x$ and $\gamma_y$ introduce non-hermiticity in the Hamiltonian such that $H_{\rm TI}(\vect{k}) \neq H^\dagger_{\rm TI}(\vect{k})$. The hopping (spin-orbit coupling) amplitudes are given by $t_{x,y}$ ($\lambda_{x,y}$). Here, $m_0$ and $\mu$ account for the crystal field splitting and chemical potential, respectively. Notice that, $H^{\rm H}_{\rm TI} (\vect {k})$ respects TRS: ${\mathcal{T}} H^{\rm H}_{\rm TI} (\vect {k}) {\mathcal{T}}^{-1}=H^{\rm H}_{\rm TI} (-\vect {k})$ and PHS: ${\mathcal{C}} H^{\rm H}_{\rm TI} (\vect {k}) {\mathcal{C}}^{-1}=-H^{\rm H}_{\rm TI} (-\vect {k})$; with $\mathcal{T}=i U_{\mathcal{T}} \mathcal{K}$ and $\mathcal{C}= U_{\mathcal{C}} \mathcal{K}$. 
The Hamiltonian (\ref{eq1_TSC}) thus takes the following compact form $\mathcal{H}(\vect{k})=\vect{N} \cdot \vect{\Gamma}$; where, $\vect{N}=\{\lambda_x \sin k_x +i \gamma_x,\lambda_y \sin k_y +i \gamma_y,m_0 -t_x \cos k_x - t_y \cos k_y,\Delta(\vect{k})\}$, $\vect{\Gamma}= \{\tau_z \sigma_x s_z,\tau_z \sigma_y s_0,\tau_z \sigma_z s_0,\tau_x \sigma_0 s_0 \}$ with  the Pauli matrices  ${\boldsymbol \tau}$, ${\boldsymbol \sigma}$, and ${\boldsymbol s}$ act on PH $(e,h)$, orbital $(\alpha,\beta)$, and spin $(\uparrow, \downarrow)$  degrees of freedom, respectively. Note that, $\mathcal{H}(\vect{k})$ obeys TRS  and PHS$^\dagger$, generated by ${\tilde U}_{\mathcal{T}}= \tau_0 \sigma_0 s_y$ and  ${\tilde U}_{\mathcal{C}}= \tau_y \sigma_0 s_y$, respectively. In addition,  $\mathcal{H}(\vect{k})$ preserves sublattice/ chiral symmetry $\mathcal{S}=\tau_y \sigma_0 s_0$ such that $\mathcal{S} \mathcal{H}(\vect{k}) \mathcal{S}^{-1}= -\mathcal{H}(\vect{k})$. Now coming to the crystalline symmetries of the model with $t_x=t_y$, $\lambda_x=\lambda_y$, and $|\gamma_x|=|\gamma_y|\ne 0$, we find that $\mathcal{H}(\vect{k})$ breaks four-fold rotation with respect to $z$, $C_4=\tau_z e^{-\frac{i \pi}{4}\sigma_z s_z}$, mirror-reflection along $x$, $\mathcal{M}_x= \tau_x \sigma_x s_0$ and mirror-reflection along $y$, $\mathcal{M}_y= \tau_x \sigma_y s_0$. As a result, $\mathcal{H}(\vect{k})$ preserves mirror-rotation I  $\mathcal{M}_{xy}=C_4 \mathcal{M}_y$ for $ \gamma_x= \gamma_y\ne 0$, and mirror-rotation II  $\mathcal{M}_{x\bar{y}}=C_4 \mathcal{M}_x$ for $\pm \gamma_x=\mp \gamma_y\ne 0$, such that $\mathcal{M}_{xy} \mathcal{H}(k_x,k_y) \mathcal{M}_{xy}^{-1}=\mathcal{H}(k_y,k_x)$ and  $\mathcal{M}_{x\bar{y}} \mathcal{H}(k_x,k_y) \mathcal{M}_{x\bar{y}}^{-1}= \mathcal{H}(-k_y,-k_x)$, respectively (see supplemental material~\cite{supp}).


We note at the outset that the definition of Majorana for NH system is different from its Hermitian analogue. The PHS$^{\dagger}$ in NH case allows us to  define a modified Hermitian conjugate operation
such that MZMs obey an effective Hermiticity ${\Gamma}^a_n=c_n + \bar{c}_n$, ${\Gamma}^b_n=i(c_n -\bar{c}_n)$,
and $\bar{\Gamma}^{a,b}_n=\Gamma^{a,b}_n$ \cite{OkumaPRL2019}; $(\bar{c}_n,c_n)$ denote the creation and annihilation operators of the Bogoliubov quasi-particles where $\bar{c}_n$ does not correspond to  Hermitian conjugate of $c_n$ in presence of  non-Hermiticity. However, the extraction of 
real MZMs individually remains unaddressed out of more than two Majorana corner modes.

The Hermitian system $\mathcal{H}^{\rm H}(\vect{k})$ hosts zero-energy Majorana corner modes, protected by the TRS,  in SOTSC phase for $m_0<\lvert t_x+t_y \rvert$ while trivially gapped for  $m_0>\lvert t_x+t_y \rvert$
\cite{Yan2018}. The NH system  becomes defective at EPs provided  $\lvert E (\vect{k}_{\rm EP}) \rvert=0$ which is in a complete contrast to the Hermitian system with $E (\vect{k})=0$ at gapless point. A close inspection of Eq.~(\ref{eq1_TSC}) suggests that four-fold degenerate  
energy bands yield  $\lvert E (0,0) \rvert=0$ [$\lvert E (\pi,\pi) \rvert=0$] for $m_0^{+,\pm}= t_x+t_y\pm\sqrt{\gamma_x^2+\gamma_y^2}$ [$m_0^{-,\pm}= -t_x-t_y\pm\sqrt{\gamma_x^2+\gamma_y^2}$]. As a result, 
the gapless phase boundaries $m^{s}_0=s (t_x+t_y)$ for the Hermitian case 
are modified in the present NH case with $m_0^{s,\pm}= s(t_x+t_y)\pm\sqrt{\gamma_x^2+\gamma_y^2}$; where, $s=\pm$ (see black lines in Figs.~\ref{Fig:sotscloop}~(a)-(c)). This refers to the fact that ${\rm Re}[E(\vect{k})]$ is gapless for $m_0^{\pm,-}< m_0  <m_0^{\pm,+}$ (see yellow-shaded region in Figs.~\ref{Fig:sotscloop}~(a)-(c)). Furthermore, $\mathcal{H}(\vect{k})$ is expected to be gapped in real sector of energy for $m_0^{-,+}< m_0  <m_0^{+,-}$, hosting NH SOTSC phase.

The above conjecture, based on periodic boundary condition (PBC), is drastically modified when the NH system (\ref{eq1_TSC}) is investigated under OBC. We show $|E|$, ${\rm Re}[E]$ and ${\rm Im}[E]$ under OBC with blue points in  Figs.~\ref{Fig:sotscloop}~(a), (b), and (c), respectively. Surprisingly, the MZMs continue to survive inside the yellow-shaded region i.e., beyond $m_0=m_0^{-,+}$ and $m_0=m_0^{+,-}$, till $m_0<\rvert \tilde{m}_0 \rvert =t_x+t_y+{\gamma_x^2}/{2\lambda_x^2}+{\gamma_y^2}/{2\lambda_y^2}$, depicted by the red line, where the ${\rm Re}[E]$ becomes gapless. All together this suggests the break-down of conventional BBC due to the non-Bloch nature of the NH Hamiltonian~\cite{YaoPRL2018,Koch2020,ZirnsteinPRL2021,TakaneJPS2021}. This apparent ambiguity in BBC affects the calculation of topological invariants, which we investigate  below.

Fig.~\ref{Fig:sotsc}~(a) demonstrates the complex-energy profile ${\rm Im}[E]$ vs ${\rm Re}[E]$ of Hamiltonian (\ref{eq1_TSC}) in real space for $m_0< \rvert \tilde{m}_0 \rvert$. We find the line gap for the NH system irrespective of the boundary conditions as the complex-energy bands do not cross a reference line in the complex-energy plane. The origin, marked with red dot in Fig.~\ref{Fig:sotsc}~(a), indicates the MZMs under OBC that are further shown the by the eight mid-gap states in ${\rm Re}[E]$-$m$ (state index) plot (see Fig.~\ref{Fig:sotsc}~(b)). Analyzing the local density of states~(LDOS) of the above MZMs, we find  sharp localization only at one corner out of the four corners \cite{NoriNHPRL2019} (see Fig.~\ref{Fig:sotsc}~(c)). 
This is a consequence of the mirror-rotation symmetries $\mathcal{M}_{xy}$ or $\mathcal{M}_{x\bar{y}}$ even though MZMs spatially coincide.  There might be additional protection from the bulk modes coming due to the emergent short-range nature of the superconducting gap \cite{Viyuela16}.
The MZMs are localized over more than a single corner when $\mathcal{M}_{xy}$ or $\mathcal{M}_{x\bar{y}}$ is broken~\cite{supp}. The MZMs are also found to be robust against onsite disorder that respects mirror-rotation and chiral symmetries~(see supplemental material~\cite{supp}). 
In addition, we remarkably find that the LDOS of the bulk modes also exhibits substantial corner localization
as depicted in the insets of Fig.~\ref{Fig:sotsc}(c) \cite{YaoPRL2018,Kunst18,helbig2020generalized}.
The above features, reflecting  the non-Block nature of the system, are referred to as the NH skin effect \cite{YaoPRL2018,Kunst18}. This is in contrast to the 
Hermitian case where only the zero-energy modes can populate four corners of the 2D square lattice \cite{schindler2018,Nag19,Ghosh2020}.

\textcolor{blue}{\textit{Topological characterization.---}}
To this end, in order to compute the topological invariant from ${\mathcal H}(\vect{k})$ characterizing the SOTSC phase under OBC, we exploit the non-Bloch nature. We need to use the complex wave-vectors to describe open-boundary eigenstates such that $\vect{k}\rightarrow \vect{k'}+i \vect{\beta}$ with $\beta_i={ \gamma_i}/{\lambda_i}$~($i=x,y$)~\cite{YaoPRLSecond2018}. Upon replacing $k_{x,y} \rightarrow k_{x,y}' -{i \gamma_{x,y}}/{\lambda_{x,y}}$, 
the renormalized topological mass $m'_0$ acquires the following form in the limit $k_{x,y}\to 0$ and $\gamma_{x,y} \to 0$
\begin{eqnarray}\label{modifiedmass}
m'_0=m_0-t_x-t_y-\frac{\gamma_x^2}{2\lambda_x^2}-\frac{\gamma_y^2}{2\lambda_y^2} \ .
\end{eqnarray}
Note that $|{\tilde m_0}|=|m'_0-m_0|$ denotes phase boundary of the SOTSC phase as obtained from Fig.~\ref{Fig:sotscloop} (b). Employing $\vect{k'} \rightarrow \vect{k'}$  in $\mathcal{H}(\vect{k})$ i.e., $\mathcal{H}(\vect{k}) \to \mathcal{H}'(\vect{k'})$, we construct the Wilson loop operator  as~\cite{benalcazarprb2017,ghosh2022dynamical}
\begin{equation}\label{wloopx}
W_{x,\vect{k'}}= F_{x,\vect{k'}+(L_x-1)\Delta_x \vect{e}_x}(t) \cdots F_{x,\vect{k'}+\Delta_x \vect{e}_x} F_{x,\vect{k'}} \ 
\end{equation}
from the non-Bloch NH Hamiltonian $\mathcal{H}'(\vect{k'})$ \cite{Huitao18,Kenta11}.
We  define $\left[F_{x,\vect{k'}}\right]_{mn}=\braket{ \Psi^{\rm L}_m (\vect{k'}+\Delta_x \vect{e}_x) | \Psi^{\rm R}_n (\vect{k'})}$, where $\ket{\Psi^{\rm R}_m(\vect{k'}) }$ ($\bra{\Psi^{\rm L}_m(\vect{k'}) }$) represents the occupied right (left) eigenvectors of the 
Hamiltonian $\mathcal{H}'(\vect{k'})$ such that ${\rm Re}[E'_m(\vect{k'})]<0$; $\Delta_i=2 \pi / L_i$, with $L_i$ being the the number of discrete points considered along $i$-th direction and $\vect{e}_i$ being the unit vector along the said direction. Notice that, the bi-orthogonalization guarantees the following
$\sum_n\ket{\Psi^{\rm R}_n(\vect{k'}) } \bra{\Psi^{\rm L}_n(\vect{k'}) }=\mathbb{I}$ and 
$\langle \Psi^{\rm L}_n(\vect{k'})| \Psi^{\rm R}_n(\vect{k'})\rangle=\delta_{mn}$; where, $n$ runs over all the energy levels irrespective of their occupations.
The  first-order polarization $\nu_{x, \mu}(k_y')$ is obtained from the  eigenvalue equation for $W_{x,\vect{k'}}$ as follows
\begin{equation}
W_{x,\vect{k'}} \ket{\nu^{\rm R}_{x, \mu}(\vect{k'})} = e^{-2 \pi i \nu_{x, \mu}(k_y')}\ket{\nu^{\rm R}_{x, \mu}(\vect{k'})} \ .
\end{equation}

Note that unlike the Hermitian case, here  $W_{x,\vect{k'}} $ is no longer unitary resulting in 
$\nu_{x, \mu}(k_y')$ to be a
complex number~\cite{LuoGainLossPRL2019}. Importantly, 
$\ket{\nu^{\rm R}_{x , \mu} (\vect{k'}) }$ ($\bra{\nu^{\rm L}_{x , \mu} (\vect{k'}) }$) designates bi-orthogonalized right (left) eigenvector  of $W_{x,\vect{k'}}$ associated with $\mu=1,\cdots,4$-th eigenvalue. 
For a (second-order topological)~SOT system, the real part of first-order polarization exhibits a finite gap in spectra such that it can be divided into two sectors as $\pm \nu_{x}(k_y')$ where each sector is two-fold degenerate. Such a structure of 
Wannier centres in  the non-Bloch case might be
relied on the mirror symmetry of the underlying Hermitian Hamiltonian \cite{benalcazarprb2017,ghosh2022dynamical}.
In order to characterize the SOT phase, we  calculate the polarization along the perpendicular $y$-direction by projecting onto each $\pm \nu_{x}$ branch. This allows us to  employ  the nested Wilson loop as follows \cite{benalcazarprb2017,ghosh2022dynamical}
\begin{equation}
W_{y,\vect{k'}}^{\pm \nu_{x}}= F_{y,\vect{k'}+(L_y-1)\Delta_y \vect{e}_y}^{\pm \nu_{x}} \cdots F_{y,\vect{k'}+\Delta_y \vect{e}_y}^{\pm \nu_{x}} F_{y,\vect{k'}}^{\pm \nu_{x}} \ .
\end{equation}
Here, $\left[F_{y,\vect{k'}}^{\pm \nu_{x}} \right]_{\mu_1 \mu_2} = \sum_{m n} \left[\nu^{\rm L}_{x , \mu_1} (\vect{k'}+\Delta_y \vect{e}_y) \right]^*_m  \left[F_{y,\vect{k'}}\right]_{mn}$ $ \left[\nu^{\rm R}_{x , \mu_2} (\vect{k'}) \right]_n$
with $\left[F_{y,\vect{k'}}\right]_{mn}=\braket{ \Psi^{\rm L}_m (\vect{k'}+\Delta_y \vect{e}_y) | \Psi^{R}_n (\vect{k'})}$. The indices $\mu_{1,2}\in \pm \nu_x$ run over the projected eigenvectors of $W_{x,\vect{k'}}$ only. We evaluate $W_{y,\vect{k'}}^{\pm \nu_{x}}$ for a given value of $k'_x$ that is the base point while calculating $W_{x,\vect{k'}}$ (\ref{wloopx}).

The nested polarization $\nu_{y, \mu'}^{\pm \nu_{x}}(k_{x}')$ can be extracted by solving  the eigenvalue equation for $W_{y,\vect{k'}}^{\pm \nu_{x}}$ 
\begin{eqnarray}
W_{y,\vect{k'}}^{\pm \nu_{x}}  \ket{\nu_{y , \mu'}^{{\rm R},\pm \nu_{x}}(\vect{k'})} = e^{-2 \pi i \nu_{y, \mu'}^{\pm \nu_{x}}(k_{x}')}\ket{\nu_{y, \mu'}^{{\rm R},\pm \nu_{x}}(\vect{k'})} \ .
\end{eqnarray}   
The average nested Wannier sector polarization $\langle \nu_{y, \mu'}^{\pm \nu_{x}}\rangle$ for the $\mu^{\prime}$-th branch, characterizing the 2D SOTSC,  is given by
\begin{equation}\label{npola}
\langle \nu_{y, \mu'}^{\pm \nu_{x}}\rangle = \frac{1}{L_x} \sum_{k_x'} {\rm Re} \left[ \nu_{y, \mu'}^{\pm \nu_{x}}(k_x') \right] \ .
\end{equation}

We explore the SOT phase diagram by investigating   mod($\langle \nu_{y, \mu'}^{\pm \nu_{x}}\rangle,1.0$) in the $m_0 \mhyphen \gamma$ ($\gamma_x=\gamma_y=\gamma$) plane keeping $t_x=t_y=\lambda_x=\lambda_y=1$ (see Fig.~\ref{Fig:sotscloop}(d)). The blue (brown) region indicates the SOTSC and trivial phase.  The  green line in Fig.~\ref{Fig:sotscloop} (d), separating the above two phases, represents the phase boundary $\tilde{m}_0=2+\gamma^2$ as demonstrated in Eq.~(\ref{modifiedmass}). On the other hand, 
the phase boundaries, obtained from bulk Hamiltonian (\ref{eq1_TSC}), are found to be  
$m_0^{+,\pm}=2 \pm \sqrt{2} \gamma$ that are depicted by 
yellow lines in Fig.~\ref{Fig:sotscloop} (d). Therefore, 
the topological invariant, computed using the non-Bloch Hamiltonian $\mathcal{H}'(\vect{k'})$, can accurately 
predict the MZMs as obtained from the real space Hamiltonian under OBC (see Fig.~\ref{Fig:sotscloop} (c)). This correspondence for very higher values of $\gamma$ no longer remains appropriate due to the possible break down of Eq.~(\ref{modifiedmass}). Even though $M_{x,y}$ are broken,  $\langle \nu_{y, \mu'}^{\pm \nu_{x}}\rangle$ ($\langle \nu_{x, \mu'}^{\pm \nu_{y}}\rangle$) yields half-integer quantization  provided $\mathcal{M}_{xy}$ ($\mathcal{M}_{x\bar{y}}$) is preserved.
Note that based on mirror
rotation and sublattice symmetries, the NH SOTSC can be shown to exhibit integer quantization in winding number similar to NH SOTI  \cite{NoriNHPRL2019} (see supplemental material~\cite{supp}).


\textcolor{blue}{\textit{Floquet generation of NH SOTSC.---}} 
Having studied the static NH SOTSC, we seek the answer to engineer dynamic NH SOTSC out of trivial phase by periodically kicking 
the on-site mass term of the Hamiltonian $\mathcal{H}(\vect{k})$ (Eq.~(\ref{eq1_TSC})) as~\cite{ghosh2020floquet2,ghosh2022systematic}
\begin{eqnarray}
m(t)=m_1 \sum_{r=-\infty}^{\infty} \delta \left(t-r T\right) \ .
\end{eqnarray}
Here, $m_1$ and $T$ represent the strength of the drive and the time-period, respectively. The Floquet operator is  formulated to be 
\begin{eqnarray}\label{Foperator}
U(\vect{k},T)&=& {\rm TO} \exp \left[-i \int_{0}^{T} dt \left\{ \mathcal{H}(\vect{k}) + m(t) \Gamma_3 \right\} \right] \non \\
&=& \exp\left[-i \mathcal{H}(\vect{k}) T \right]  \exp\left[-i m_1 \textcolor{red}{\Gamma_3} \right]  \ , 
\end{eqnarray}
where TO denotes the time ordering. Notice that $m_0 >|t_x +t_y+ \sqrt{\gamma_x^2 + \gamma_y^2}|$ such that the underlying static NH Hamiltonian  $\mathcal{H}(\vect{k})$ remains in the trivially gapped phase.
Having constructed the Floquet operator $U(\vect{k},T)$, we resort to OBCs and diagonalize the Floquet operator to obtain the quasi-energy spectrum for the system. 
We depict the real part of the quasi-energy $\mu_m$ as function of the state index $m$ in Fig.~\ref{Fig:Floquetsotsc} (a)  where frequency of the drive is  higher than the band-width of the system.
The existence of eight MZMs is a signature of the NH Floquet SOTSC phase. The  LDOS for the MZMs displays substantial  localization only at one corner in Fig.~\ref{Fig:Floquetsotsc}(b). Insets show the NH skin effect where the bulk modes at finite energy also have 
a fair amount of corner localization. 

In order to topologically characterize the above MZMs, we again make use of the non-Bloch form. Instead of the static Hamiltonian, we derive the high-frequency effective Floquet Hamiltonian, in the limit $T \rightarrow 0$ and  $m_1 \rightarrow 0$ to analyze the situation   
\begin{eqnarray}
H_{\rm Flq}(\vect{k})&\approx& \mathcal{H} (\vect{k}) +\frac{m_1}{T} \Gamma_1 +m_1 \sum_{j=2}^{4}N_j  \Gamma_{j1} \ ,
\end{eqnarray}
with $\Gamma_{21}=-\sigma_y s_z, \Gamma_{31}=\sigma_x, \Gamma_{41}=-\tau_y\sigma_z$. Upon substitution of $\vect{k}\rightarrow \vect{k'}+i \vect{\beta}$, the modified mass term in $H'_{\rm Flq}(\vect{k'})$
reads as
\begin{eqnarray}\label{modifiedmassFloquet}
m'_0=m_0-t_x-t_y-\frac{\gamma_x^2}{2\lambda_x^2}-\frac{\gamma_y^2}{2\lambda_y^2}+\frac{m_1}{T} \ .
\end{eqnarray}
Evaluating the effective Floquet nested Wannier sector polarization $\langle \nu_{y, \mu'}^{{\rm F}, \pm \nu_x} \rangle$ numerically from non-Bloch Floquet operator $U'(\vect{k'},T)$ \cite{ghosh2022dynamical,Huang2020},
we obtain the Floquet phase diagram in $m_1$-$\gamma$ plane as shown in Fig.~\ref{Fig:Floquetsotsc} (c).
The non-Bloch Floquet operator can be considered as the dynamic analog of the  non-Bloch NH Hamiltonian $H'(\vect{k'})$. In particular, we use bi-orthogonalized
$\ket{\Psi^{\rm R}_{\rm F}(\vect{k'}) }$ ($\bra{\Psi^{\rm L}_{\rm F}(\vect{k'}) }$), representing the occupied right (left) quasi-states of $U'(\vect{k'},T)$ with quasi-energy $-\pi/T<{\rm Re}[\mu]<0$, to construct the Wilson loops  $W^{\rm F}_{x,\vect{k'}}$ for the driven case. Following identical line of arguments, presented for the static case,  $W_{y,\vect{k'}}^{{\rm F},\pm \nu_{x}}$ is obtained from  
$\left[F_{y,\vect{k'}}^{{\rm F},\pm \nu_{x}} \right]_{\mu_1 \mu_2} = \sum_{m n} \left[\nu^{{\rm F},\rm L}_{x , \mu_1} (\vect{k'}+\Delta_y \vect{e}_y) \right]^*_m  \left[F^{\rm F}_{y,\vect{k'}}\right]_{mn}$ $ \left[\nu^{{\rm F},\rm R}_{x , \mu_2} (\vect{k'}) \right]_n$
with $\left[F^{\rm F}_{y,\vect{k'}}\right]_{mn}=\braket{ \Psi^{\rm L}_{{\rm F}m} (\vect{k'}+\Delta_y \vect{e}_y) | \Psi^{R}_{{\rm F}n} (\vect{k'})}$ and 
$\ket{\nu^{\rm F,R}_{x , \mu} (\vect{k'}) }$ ($\bra{\nu^{\rm F,L}_{x , \mu} (\vect{k'}) }$) designates bi-orthogonalized right (left) eigenvector  of $W^{\rm F}_{x,\vect{k'}}$. 
Interestingly, this is similar   to the static phase diagram where the phase boundary is accurately explained by Eq.~(\ref{modifiedmassFloquet}).

We further analyze the problem for lower frequency regime to look for anomalous Floquet modes at quasienergy ${\rm Re}[\mu] =\pm \pi$~\cite{ghosh2022systematic,ghosh2022dynamical}. We depict one such scenario for $\Omega=2\pi/T=5.0$ in Fig.~\ref{Fig:anomaloussotsc} (a) where eight anomalous $\pi$-modes appear simultaneously with regular eight $0$-modes. The corresponding LDOS for $0$-mode and $\pi$-mode are shown in Fig.~\ref{Fig:anomaloussotsc}(b) and (c), respectively. Interestingly, the $0$-mode and the $\pi$-mode populate  different corners of the system. As a signature of NH skin effect, we show 
the LDOS for two bulk states in the inset I1 and I2 of Fig.~\ref{Fig:anomaloussotsc}(b). The localization profile of the zero-energy states and bulk states are unique to the NH system that can not be explored in its Hermitian counterpart.

\textcolor{blue}{\textit{Discussions.---}}
The number of MZMs can be tuned in our case by the application of magnetic field similar to the Hermitian SOTSC phase \cite{ghosh2020floquet2}. The long-range hopping provides another route to enhance the number of MZMs that can in principle be applicable for the non-Hermitian case as well \cite{DeGottardi13,Benalcazar22}. Interestingly, Floquet driving delivers an alternative handle to generate long-range hopping effectively out of the short-range NH model such that the number of MZMs are varied (see supplemental material~\cite{supp}). 
Interestingly, Hermitian and non-Hermitian phases belong to the Dirac and non-Hermitian Dirac universality classes \cite{Wei17,Arouca20}. In the case of HOT phases,  one expects different critical exponents with respect to
the usual Dirac model. The breakdown of BBC and skin effect are intimately related to such non-Hermitian Dirac universality class. The edge theory, computed from Hermitian HOT model, is modified due to the non-Hermiticity with the possible non-Bloch form. 
Given the experimental realization of spin-orbit coupling \cite{huang2016experimental,wu2016realization}, non-Hermiticity \cite{li2019observation,Wange19} and theoretical  proposals on topological superfluidity \cite{Jia19,buhler2014majorana} in optical lattice, we believe that the cold atom
systems might be a suitable platform for the potential experimental realization of our findings  
\cite{Makris08,Miyake13,mckay2011cooling}. However, we note that    the superconductivity  might be hard to achieve in the NH setting. 

\textcolor{blue}{\textit{Summary and conclusions.---}}
In this article, we  consider 2D NH TI, proximized with $d$-wave superconductivity, to investigate the emergence of NH SOTSC phase. From the analysis of  EPs on the bulk NH Hamiltonian under PBC, one can estimate the gapped and gapless phase in terms of the real energies (see Fig.~\ref{Fig:sotscloop}). By contrast, the MZMs, obtained from the real space NH Hamiltonian under OBC, do not immediately vanish inside the bulk gapless region (see Fig.~\ref{Fig:sotsc}). This apparent breakdown of the BBC can be explained by the non-Bloch nature of the NH Hamiltonian that further results in  the MZMs residing at only one corner while the bulk modes  populate the  boundaries. While the later is dubbed as NH skin effect. We propose the  nested polarization for topologically characterize the MZMs upon exploiting the non-Bloch form of the  complex wave vectors.  This resolves the anomaly between the phase boundaries, obtained from OBC and PBC, in the topological phase diagram.   
Finally, we adopt a mass-kick drive to illustrate the Floquet generation of NH SOTSC out of the trivial phase and characterize it using non-Bloch Floquet nested Wannier sector polarization (see Fig.~\ref{Fig:Floquetsotsc}). In addition, we demonstrate the emergence of  anomalous $\pi$-mode following such drive when the frequency is lowered (see Fig.~\ref{Fig:anomaloussotsc}). The mirror symmetries $M_{x,y}$ play crucial role in characterizing the anomalous $\pi$-modes \cite{Huang2020,ghosh2022dynamical}. Therefore, such characterization
in the absence of mirror symmetries is a future problem.

\textit{Acknowledgments.---} A.K.G. acknowledges SAMKHYA: High-Performance Computing Facility provided by Institute of Physics, Bhubaneswar, for numerical computations. We thank Arijit Saha for useful discussions. 

\bibliography{bibfile}{}


\clearpage
\begin{onecolumngrid}
	\begin{center}
		{\fontsize{12}{12}\selectfont
			\textbf{Supplemental Material for ``Non-Hermitian higher-order topological superconductors in two-dimension: statics and dynamics''\\[5mm]}}
		{\normalsize Arnob Kumar Ghosh\orcidA{},$^{1,2}$, and Tanay Nag\orcidB{},$^{3}$\\[1mm]}
		{\small $^1$\textit{Institute of Physics, Sachivalaya Marg, Bhubaneswar-751005, India}\\[0.5mm]}
		{\small $^2$\textit{Homi Bhabha National Institute, Training School Complex, Anushakti Nagar, Mumbai 400094, India}\\[0.5mm]}
		{\small $^3$\textit{Department of Physics and Astronomy, Uppsala University, Box 516, 75120 Uppsala, Sweden}\\[0.5mm]}
		{}
	\end{center}
	
	\normalsize
	\newcounter{defcounter}
	\setcounter{defcounter}{0}
	\setcounter{equation}{0}
	\renewcommand{\theequation}{S\arabic{equation}}
	\setcounter{figure}{0}
	\renewcommand{\thefigure}{S\arabic{figure}}
	\setcounter{page}{1}
	\pagenumbering{roman}
	
	\renewcommand{\thesection}{S\arabic{section}}

	\section{Underlying spatial symmetry}
	\label{Sec:symmetry}
	In this section, we analyze the crystalline symmetries associated with the non-Hermitian (NH) second-order topological superconductor (SOTSC) as presented in Eq. (1) of the main text.  
	In order to understand the spatial symmetry of the model, we first analyze the underlying Hermitian version of our model $\mathcal{H}^{\rm H}(\vect{k})= \lambda_x \sin k_x \tau_z \sigma_x s_z + \lambda_y \sin k_y \tau_z \sigma_y s_0 + (m_0 -t_x \cos k_x - t_y \cos k_y) \tau_z \sigma_z s_0  + \Delta (\cos k_x - \cos k_y) \tau_x \sigma_0 s_0 $ as demonstrated in Eq. (1) of the main text. The Hermitian Hamiltonian obeys the following spatial symmetries- 
	
	\textcolor{black}{
		\begin{itemize}
			\item mirror symmetry along $x$ with $\mathcal{M}_x= \tau_x \sigma_x s_0$: $\mathcal{M}_x \mathcal{H}^{\rm H}(k_x,k_y) \mathcal{M}_x^{-1}= \mathcal{H}^{\rm H}(-k_x,k_y) $, 
			\item mirror symmetry along $y$ with $\mathcal{M}_y= \tau_x \sigma_y s_0$: $\mathcal{M}_y \mathcal{H}^{\rm H}(k_x,k_y) \mathcal{M}_y^{-1}= \mathcal{H}^{\rm H}(k_x,-k_y) $,
			\item four-fold rotation with $C_4= \tau_z e^{-\frac{i \pi}{4}\sigma_z s_z}$: $C_4 \mathcal{H}^{\rm H}(k_x,k_y)C_4^{-1}= \mathcal{H}^{\rm H}(-k_y,k_x) $,  if $t_x=t_y$ and $\lambda_x=\lambda_y$,  
			\item mirror rotation I with $\mathcal{M}_{xy}=C_4 \mathcal{M}_y$: $\mathcal{M}_{xy} \mathcal{H}^{\rm H}(k_x,k_y) \mathcal{M}_{xy}^{-1}= \mathcal{H}^{\rm H}(k_y,k_x)$, if $t_x=t_y$ and $\lambda_x=\lambda_y$,
			\item mirror rotation II with $\mathcal{M}_{x\bar{y}}=C_4 \mathcal{M}_x$: $\mathcal{M}_{x\bar{y}} \mathcal{H}^{\rm H}(k_x,k_y) \mathcal{M}_{x\bar{y}}^{-1}= \mathcal{H}^{\rm H}(-k_y,-k_x)$, if $t_x=t_y$ and $\lambda_x=\lambda_y$, 
			\item sublattice/ chiral symmetry with $\mathcal{S}=\tau_y \sigma_0 s_0$:  $\mathcal{S} \mathcal{H}^{\rm H}(\vect{k}) \mathcal{S}^{-1}= -\mathcal{H}^{\rm H}(\vect{k}) $.
		\end{itemize}
	}
	
	The extended Hermitian Hamiltonian $\tilde{\mathcal{H}}(\vect{k})$ is found to be crucial for analyzing the crystalline symmetries of the NH Hamiltonian that  can be wriiten as follows  \cite{Kawabata20b}
	\begin{equation}
		\tilde{\mathcal{H}}(\vect{k})=
		\begin{pmatrix}
			0 & \mathcal{H}(\vect{k}) \\
			\mathcal{H}(\vect{k})^{\dagger} & 0\   \\
		\end{pmatrix}\ .
		\label{eq1_TSC}
	\end{equation}
	To be precise, $\tilde{\mathcal{H}}(\vect{k})=\lambda_x \sin k_x \mu_x \tau_z \sigma_x s_z + \gamma_x \mu_y \tau_z \sigma_x s_z+ \lambda_y \sin k_y \mu_x \tau_z \sigma_y s_0 + \gamma_y \mu_y \tau_z \sigma_y s_0  + (m_0 -t_x \cos k_x - t_y \cos k_y) \mu_x \tau_z \sigma_z s_0  + \Delta (\cos k_x - \cos k_y) \mu_x \tau_x \sigma_0 s_0 $. 
	For $\gamma_x\ne 0, \gamma_y= 0$ ($\gamma_x= 0, \gamma_y \ne 0$), $\tilde{\mathcal{H}}(\vect{k})$ breaks $\mathcal{M}_x=\mu_x\tau_x \sigma_x s_0$ ($\mathcal{M}_y=\mu_x \tau_x \sigma_y s_0$).  When $\gamma_x\ne 0, \gamma_y\ne 0$, $\tilde{\mathcal{H}}(\vect{k})$ breaks $C_4= \mu_x \tau_z e^{-\frac{i \pi}{4}\sigma_z s_z} $. For $\gamma_x=\gamma_y\ne 0$, $\tilde{\mathcal{H}}(\vect{k})$ preserves  
	$\mathcal{M}_{xy}=C_4 \mathcal{M}_y$ and $\mathcal{M}_{x\bar{y}}=C_4 \mathcal{M}_x$. On the other hand, $\tilde{\mathcal{H}}(\vect{k})$ preserves the chiral symmetry $\mathcal{S}=\mu_0 \tau_y \sigma_0 s_0$. 
	
	Therefore, the NH Hamiltonian (Eq. (1) of the main text) breaks and respects  the following symmetries. When $\gamma_x \ne 0$~($\gamma_y \neq 0$) $\mathcal{H}(\vect{k})$ breaks $\mathcal{M}_x$~($\mathcal{M}_y$) and when both $\gamma_{x,y} \ne 0$, $\mathcal{H}(\vect{k})$ breaks $C_4$ symmetry. The Hamiltonian $\mathcal{H}(\vect{k})$ obeys both $\mathcal{M}_{xy}$ and $\mathcal{M}_{x\bar{y}}$ provided $\gamma_x=\gamma_y$. $\mathcal{H}(\vect{k})$ also preserves the sublattice/ chiral symmetry $\mathcal{S}$ like its Hermitian counterpart.

	\section{Effect of asymmetric $\gamma$ and the localization of the Majorana zero-modes} \label{Sec:S1}
	In this section, we study the localization of the Majorana zero-modes~(MZMs) for different values of $\gamma_x$ and $\gamma_y$~\cite{NoriNHPRL2019}. In Fig.~2 (c) of the main text, we demonstrate that the MZMs are located only at the top-right corner of the system when $\gamma_x=\gamma_y=\gamma$. However, one can choose $\gamma_x=-\gamma_y=\gamma$~($\gamma_y=-\gamma_x=\gamma$), and following the calculation of the local density of states~(LDOS), it turns out that the MZMs are now localized at the bottom-right~[Fig.~\ref{asymmrtric}~(a)]~(top-left [Fig.~\ref{asymmrtric}~(b)]) corner of the system. When $\gamma_x=\gamma_y=-\gamma$, the MZMs are localized at the bottom-left corner of the system~[see Fig.~\ref{asymmrtric}~(c)]. However, if we break the mirror rotation symmetry $\mathcal{M}_{xy}$ by considering $\gamma_x \neq \gamma_y$, we depict that the MZMs are localized at more than one corner of the system in Fig.~\ref{asymmrtric}~(d) and (e). We obtain a non-quantized value for the topological invariant defined by employing the nested Wilson loop in the main text [Eq.(7)] when the mirror-rotation symmetry $\mathcal{M}_{xy}$ is broken. We can find that either right(left)-bottom  or left(right)-top corner is only occupied for mirror-rotation symmetry $\mathcal{M}_{x\bar{y}}$ ($\mathcal{M}_{xy}$). 
	Therefore, the localization profile of MZMs over a given corner is caused by the  interplay between the mirror-rotation symmetry and non-Hermiticity \cite{NoriNHPRL2019}.

	\begin{figure}[H]
		\centering
		\subfigure{\includegraphics[width=0.9\textwidth]{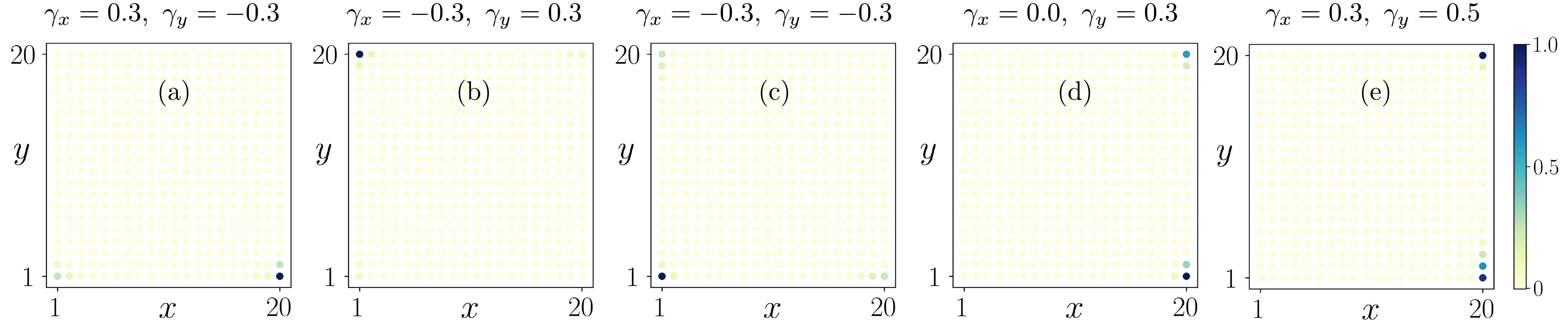}}
		\caption{We depict the local density of states for the Majorana zero-modes as a function of the system dimensions in (a) for $\gamma_x=-\gamma_y=0.3$, (b) for $\gamma_y=-\gamma_x=0.3$, (c) for $\gamma_x=\gamma_y=-0.3$, (d) for $\gamma_x=0.0,\gamma_y=0.3$, and (e) for $\gamma_x=0.3,\gamma_y=0.5$. The other parameters used here are  $t_x=t_y=\lambda_x=\lambda_y=\Delta=1.0$.
		}
		\label{asymmrtric}
	\end{figure}

	\section{Symmetry constraints for Wannier bands}
	\label{Sec:sym_wc}
	
	\begin{figure}[H]
		\centering
		\subfigure{\includegraphics[width=0.9\textwidth]{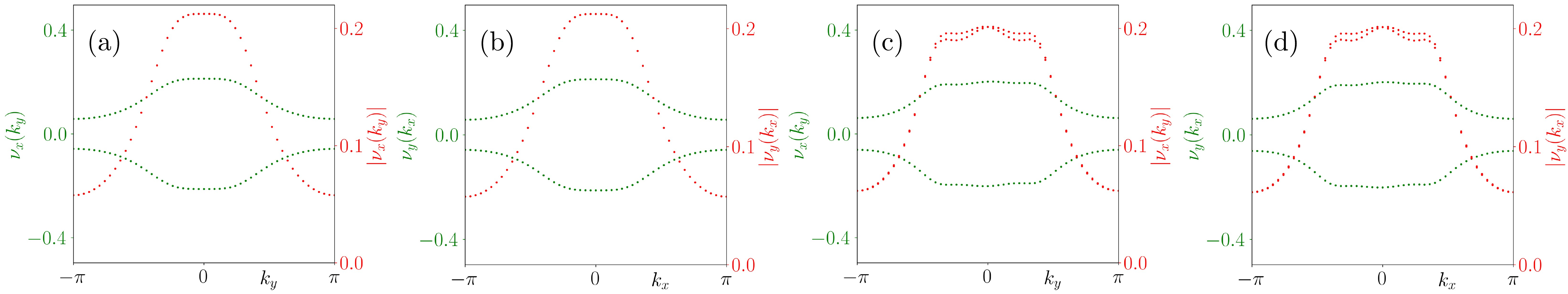}}
		\caption{We exhibit the first-order Wannier centres $\nu_x(k_y)$ (left axis)  and $\lvert \nu_x(k_y) \rvert$ (right axis) as function of $k_y$ in (a) for Hermitian case (c) for non-Hermitian case. In the non-Hermitian case the breaking of mirror symmetry $M_x$ causes the irreversible nature of first-order branches \ie $\nu_x(k_y) \not\to -\nu_x(k_y)$, which is more evident from $\lvert \nu_x(k_y) \rvert$. (b), (d) We repeat (a) and (b) for $\nu_y(k_x)$. \textcolor{black}{Note that for better visibility, we adopt the scale for $\nu_x(k_y),~\nu_y(k_x)$ to be between $[-0.5,0.5]$ while  $|\nu_x(k_y)|,~|\nu_y(k_x)|$ are bounded between $[0.0,0.22]$. }
		}
		\label{firstorder}
	\end{figure}

	In this section, we show that how the first- and second-order Wannier centres behave under the above 
	spatial symmetries. The first-order Wannier centres $\nu_x(k_y)$ and second-order Wannier centres $\nu_y^{\nu_x}$ behave in the following way \cite{benalcazarprb2017}- mirror-rotation along $x$-direction
	$M_x$: $\nu_x(k_y)\to -\nu_x(k_y)$  [see Fig.~\ref{firstorder}], and $\nu_y^{\nu_x}(k_x)\to \nu_y^{-\nu_x}(-k_x)$; $M_x$ causes the first-order branches to appear in pairs, mirror-rotation along $y$-direction.
	$M_y$: $\nu_x(k_y)\to \nu_x(-k_y)$, and $\nu_y^{\nu_x}(k_x) \to -\nu_y^{\nu_x}(k_x)$; $M_y$
	defines the shape of the first-order branches. The four-fold rotation $C_4$  and mirror rotations $\mathcal{M}_{xy}$, $\mathcal{M}_{x\bar{y}}$ interchange the branches,
	$C_4$: $\nu_x(k_y)\to -\nu_y(k_x)$, and $\nu_y^{\nu_x}(k_x) \to \nu_x^{-\nu_y}(-k_y)$, 
	$\mathcal{M}_{xy}$:  $\nu_x(k_y)\to \nu_y(k_x)$, and $\nu_y^{\nu_x}(k_x) \to \nu_x^{\nu_y}(k_y)$
	$\mathcal{M}_{x\bar{y}}$: $\nu_x(k_y)\to -\nu_y(-k_x)$, and $\nu_y^{\nu_x}(k_x) \to - \nu_x^{-\nu_y}(-k_y)$.
	In our case, the first-order branches do not follow the above relations as $M_{x,y}$ symmetries are broken.   However, the second-order polarization i.e., nested polarization mod($\langle \nu_{y, \mu'}^{\pm \nu_{x}}\rangle,1.0$) and mod($\langle \nu_{x, \mu'}^{\pm \nu_{y}}\rangle,1.0$) both are found to be $0.5$ following the mirror rotation symmetries generated by  $\mathcal{M}_{xy}$ and $\mathcal{M}_{x\bar{y}}$. Moreover, under these mirror rotation symmetries, one can show that $\nu_x^{\nu_y}(k_y) \to -\nu_x^{-\nu_y}(-k_y)$ and $\nu_y^{\nu_x}(k_x) \to -\nu_y^{-\nu_x}(-k_x)$.  
	We further notice that when $\mathcal{M}_{xy}$ is broken, $\langle \nu_{y, \mu'}^{\pm \nu_{x}}\rangle$ and $\langle \nu_{x, \mu'}^{\pm \nu_{y}}\rangle$ do not produce a quantized value of $0/0.5$. Therefore, these spatial symmetries can predict the quantization of the second-order polarization that determines the second-order topology. 
	In this regard, we would like to
	comment that the winding number,  based on the mirror rotation symmetry $\mathcal{M}_{xy}$ and the sublattice symmetry $\mathcal{S}$, is shown to topologically characterize the NH higher-order
	topological insulator (HOTI) \cite{NoriNHPRL2019}.  
	We believe that it is possible to define such a 
	winding number yielding quantized values in the SOTSC phase as our NH model preserves mirror rotation and   sublattice symmetries.

	\textcolor{black}{Note that winding number, usually characterizing a first-order topological phase provided Hamiltonian preserves sublattice symmetry \cite{Shinsei16}, can also identify a second-order topological phase while appropriately defined with other symmetry constraints. The  winding number for a Hamiltonian physically means  how many times the Hamiltonian encircles a given point when a parameter is cyclically varied. On the other hand, half integer eigenvalues of nested Wilson loop refers to the fact that Wannier centers, located at a high-symmetry point with respect to the mirror symmetries, lie half-way between the lattice sites for higher-order topological phase \cite{benalcazarprb2017}. Therefore, both the above invariant captures the non-trivial nature of the underlying wave-functions in the Brillouin zone  through either Wannier center or winding number once a certain set of symmetries is present. The winding number thus can be considered to be equivalent to our nested Wilson loop eigenvalues i.e, second-order Wannier values.  The quantization in $\langle \nu_{y, \mu'}^{\pm \nu_{x}}\rangle$ and $\langle \nu_{x, \mu'}^{\pm \nu_{y}}\rangle$ might be intimately connected to winding number as stated above.  However, the detailed mathematical proof is an open question yet, to the best of our knowledge, that we leave  for future studies.}

	
	\section{$s$-wave NH SOTSC}
	\label{Sec:s_wave}

	
	In the main text, we discuss $d$-wave proximized to engineer NH SOTSC phase. We here   illustrate that 
	NH higher-order topological superconductor (HOTSC) can be contemplated for $s$-wave superconductivity with pairing amplitude $\Delta_s$ as follows~\cite{YanPRB2019,ghosh2020floquet} 
	$\mathcal{H}(\vect{k})=\vect{N} \cdot \vect{\Gamma}$; where, $\vect{N}=\{\lambda_x \sin k_x +i \gamma_x,\lambda_y \sin k_y +i \gamma_y,m_0 -t_x \cos k_x - t_y \cos k_y,\Delta_s,\Lambda (\cos k_x -\cos k_y )  \}$, $\vect{\Gamma}= \{\tau_z \sigma_x s_z,\tau_z \sigma_y s_0,\tau_z \sigma_z s_0,\tau_x \sigma_0 s_0,\tau_0 \sigma_x s_x \}$. 
	The last term proportional to $\Lambda$ represents $C_4$ symmetry breaking Wilson-Dirac mass term. We depict the corresponding complex-energy bands
	in complex energy plane, eigenvalue spectra  highlighting the mid-gap states, and the LDOS associated to the MZMs and bulk modes in 
	Fig.~\ref{Fig:swaveSOTSC} (a), (b), and (c), respectively. The NH SOTSC phase with $\Delta_s$  can also be obtained using in-plane magnetic field instead of 
	$C_4$ symmetry breaking \cite{Wu2020,ghosh2020floquet}. The topological characterization of these phases we leave for future studies.   
	

	\begin{figure}[]
		\centering
		\subfigure{\includegraphics[width=0.68\textwidth]{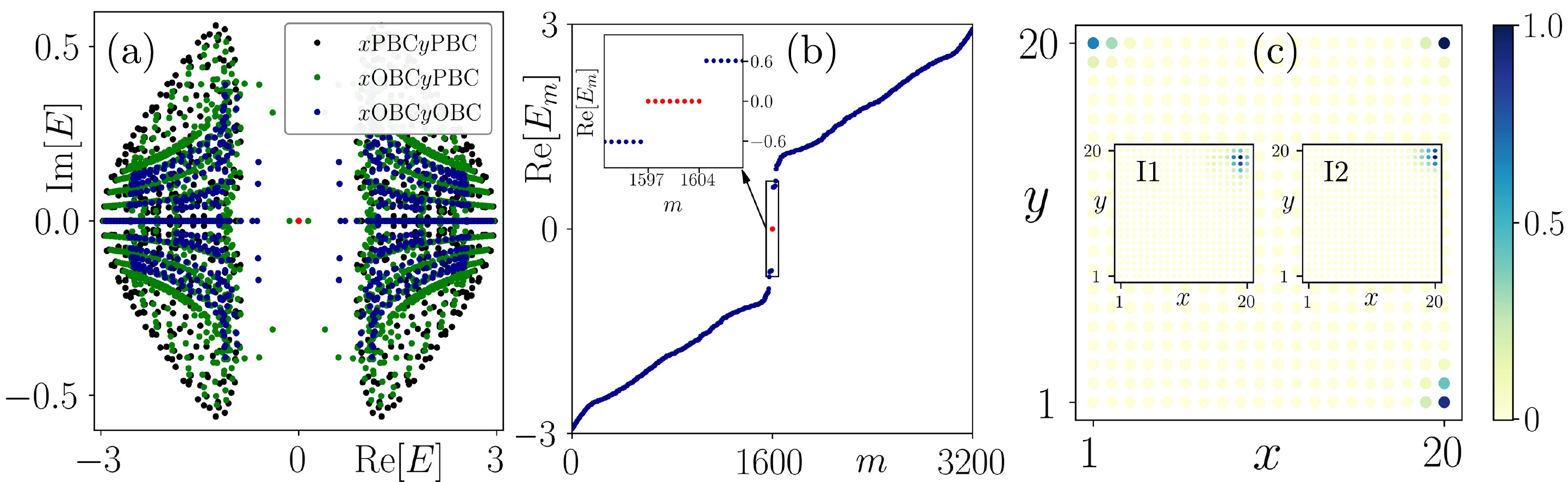}}
		\caption{(a) The  eigenvalue spectrum for the real space 2D system, obeying PBC in both direction (black dots), PBC in $y$ and OBC in $x$-direction (green dots), and OBC in both directions (blue dots) are depicted in complex energy plane. The zero-energy mode, obtained from OBC, is marked by red dots. (b) ${\rm Re}[E_m]$ as a function of the state index $m$ is displayed  where eight mid-gap MZMs are highlighted in the inset. (c) The LDOS, associated with eight MZMs in (b),   show sharp localization only at one corner. We choose  $E_m=-2.666466$ and $E_m=-1.713832+0.038500i$ for insets I1 and I2 in (c), respectively. The parameters used here are  $t_x=t_y=\lambda_x=\lambda_y=\Lambda=1.0$ and $\Delta_s=\gamma_x=\gamma_y=0.4$.
		}
		\label{Fig:swaveSOTSC}
	\end{figure}
	
	\begin{figure}[H]
		\centering
		\subfigure{\includegraphics[width=0.9\textwidth]{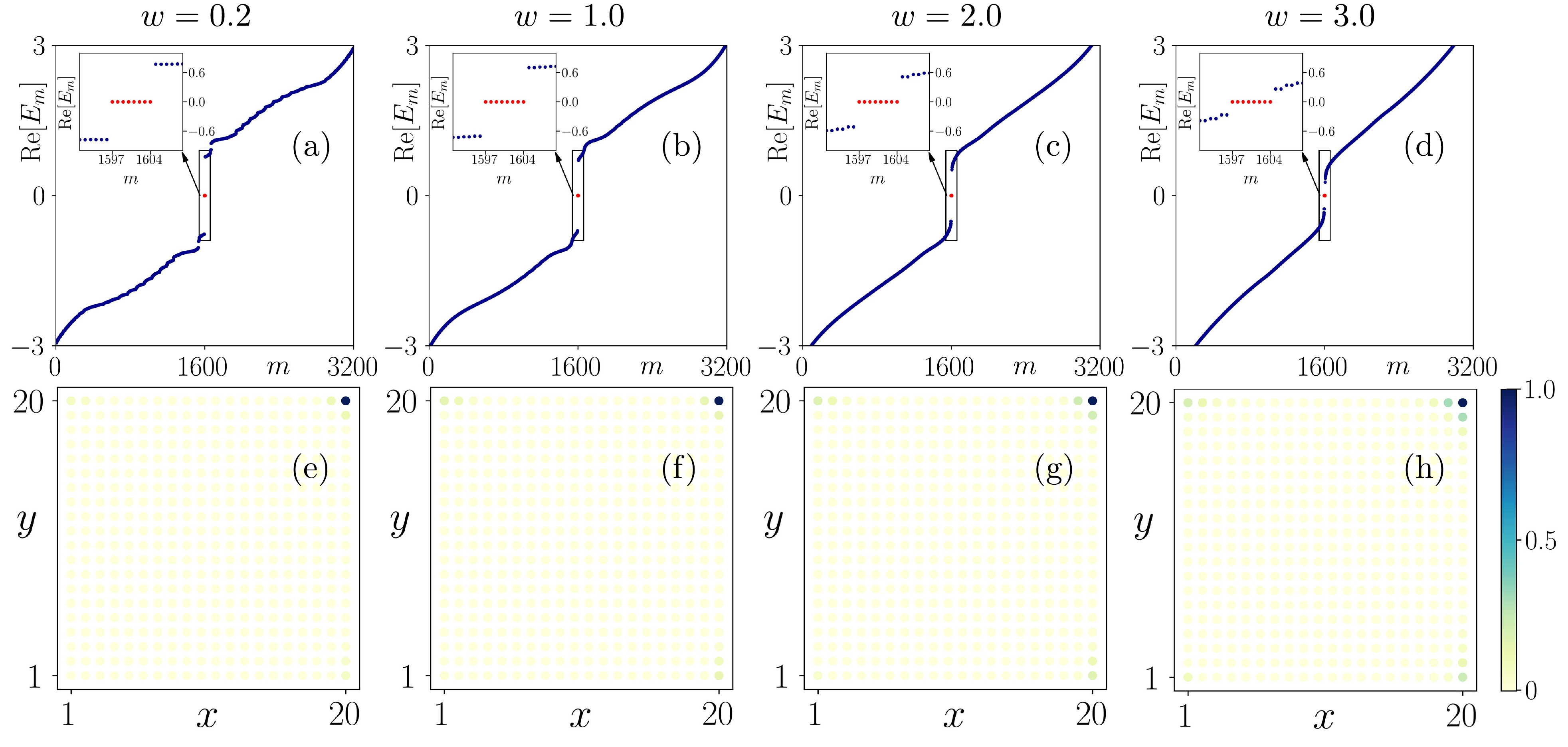}}
		\caption{We depict the eigenvalue spectra $E_m$ and local density of states in (a), (e) for $w=0.2$, (b), (f) for $w=1.0$, (c), (g) for $w=2.0$, and (d), (h) for $w=3.0$. One see that even for a very high-disorder strength, the Majorana zero modes are roboust. 
		}
		\label{onsitedisorder}
	\end{figure}

	\section{Effect of onsite disorder} \label{Sec:S5}
	In this section,  we investigate the effect of on-site disorder to investigate the  stability of the MZMs.  We consider an onsite disorder potential of the form $V(i,j)=\sum_{i,j} V_{ij} \Gamma_3$. Here, $ V_{ij}$ is randomly distributed in the range $ V_{ij} \in \left[ -\frac{w}{2}, \frac{w}{2}\right]$; while $w$ accounts for the strength of the disorder potential.  The  disordered NH Hamiltonian ${\mathcal H}'(i,j)={\mathcal H}(i,j) + V(i,j) \Gamma_3$ respects the chiral symmetry while investigating the extended Hermitian Hamiltonian $\tilde{{\mathcal H}}'$: $\mathcal{S} \tilde{{\mathcal H}}'  \mathcal{S}^{-1}= -\tilde{{\mathcal H}}' $ with $\mathcal{S}=\mu_0 \tau_y \sigma_0 s_0$. 
	We also note that on-site disorder preserves 
	$\mathcal{M}_{xy}$.
	In the presence of the onsite disorder, the real-space Hamiltonian $\mathcal{H}'$ is given as
	\begin{eqnarray}
		\mathcal{H}'&=& \sum_{i,j} \Psi_{i,j}^\dagger \left[i \gamma_x \Gamma_1 i \gamma_y \Gamma_2 + \left (m_0 + V_{ij} \right) \Gamma_3 \right] \Psi_{i,j} + \sum_{\langle i,j \rangle_x} \Psi_{i,j}^\dagger \left[-\frac{i \lambda_x}{2} \Gamma_1- \frac{t_x}{2} \Gamma_3 +\frac{\Delta}{2} \Gamma_4 \right] \Psi_{i,j} \non \\
		&&+ \sum_{\langle i,j \rangle_y} \Psi_{i,j}^\dagger \left[-\frac{i \lambda_y}{2} \Gamma_2- \frac{t_y}{2} \Gamma_3 -\frac{\Delta}{2} \Gamma_4 \right] \Psi_{i,j} \ .
	\end{eqnarray}
	Here, $\Psi_{i,j}$ is a $8 \times 1$ matrix consisting of the annihilation operator in the particle-hole, orbital, and spin subspace at a lattice site $(i,j)$. And $\langle i,j\rangle_x$~($\langle i,j\rangle_y$) represents the nearest neighbor hopping along $x$~($y$) direction. We consider 500 disorder configurations and depict the eigenstates $E_m$ as a function of the state index $m$ and the LDOS of the MZMs in Fig.~\ref{onsitedisorder} for different disorder strengths.
	The MZMs remain localized at a given corner due to the 
	mirror symmetry preserving nature of the on-site disorder.
	We can conclude from the eigenvalue spectra and the LDOS plot that the MZMs are robust and against the onsite disorder. 
	
	\section{Tuning the number of MCMs dynamically} \label{Sec:MZMs_tune}
	
	In this section, we discuss how Floquet engineering permits us to generate more in-gap states (both $0$- and $\pi$-gap) by tuning the different driving parameters suitably. The generation of multiple in-gap states is attributed to higher-order hoppings generated in the dynamical system. The mass-kick protocol, introduced in Eq.~(9) of the main text, can be used to create multiple Majorana corner modes (MCMs). We depict the real part of the quasienergy spectra as a function of the driving frequency $\Omega$ in Fig.~\ref{multiplemode}~(a). In order to understand the generation of the MCMs more clearly, we exemplify two cases in Fig.~\ref{multiplemode}~(b) and (c) for $\Omega=4.6$ and $1.9$, respectively. In Fig.~\ref{multiplemode}~(b), the number of $0$-MCMs is twice that of the static case, whereas the number of $\pi$-MCMs remains the same. While in Fig.~\ref{multiplemode}~(c), the number of $\pi$-MCMs is thrice compared to the case in the main text (see Fig. 4 (a) of the main text).
	
	\begin{figure}[H]
		\centering
		\subfigure{\includegraphics[width=0.75\textwidth]{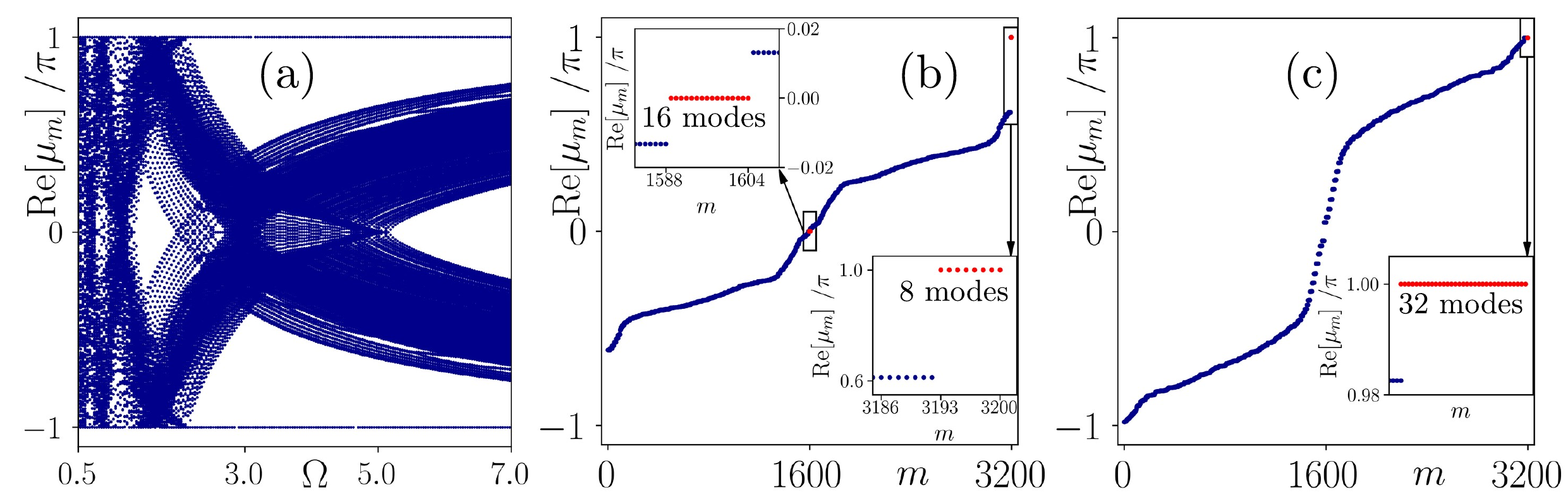}}
		\caption{We depict the real part of the quasienergy spectra Re$[\mu_m]$ as function of the driving frequency $\Omega$  in (a). We exhibit the quasienergy spectra as function of the state index $m$ for two individual cases in (b) with 16 0-MCMs and 8 $\pi$-MCMs, and (b) 32 $\pi$-MCMs. We consider $\Omega=4.6$ for (b) and $\Omega=1.9$ for (c). While all the other parameters take the value: $m_0=0.0,~m_1=2.5$, $t_x=t_y=\lambda_x=\lambda_y=\Delta=1.0$, and $\gamma_x=\gamma_y=0.4$.
		}
		\label{multiplemode}
	\end{figure}

\end{onecolumngrid}

\end{document}